\documentclass[12pt]{article}
\begin{document}
\title{\bf Teleparallel Energy-Momentum Distribution
of Lewis-Papapetrou Spacetimes}

\author{M. Sharif \thanks{msharif@math.pu.edu.pk} and M Jamil Amir
\thanks{mjamil.dgk@gmail.com}\\
Department of Mathematics, University of the Punjab,\\
Quaid-e-Azam Campus, Lahore-54590, Pakistan.}

\date{}

\maketitle

\begin{abstract}
In this paper, we find the energy-momentum distribution of
stationary axisymmetric spacetimes in the context of teleparallel
theory by using M$\ddot{o}$ller prescription. The metric under
consideration is the generalization of the Weyl metrics called the
Lewis-Papapetrou metric. The class of stationary axisymmetric
solutions of the Einstein field equations has been studied by
Galtsov to include the gravitational effect of an {\it external}
source. Such spacetimes are also astrophysically important as they
describe the exterior of a body in equilibrium. The energy density
turns out to be non-vanishing and well-defined and the momentum
becomes constant except along $\theta$-direction. It is interesting
to mention that the results reduce to the already available results
for the Weyl metrics when we take $\omega=0$.
\end{abstract}



\section{Introduction}

Localization of energy-momentum is one of the oldest and thorny
problems [1] in General Relativity (GR) which is still without any
acceptable answer in general. Being a natural field, it is expected
that gravity should have its own local energy-momentum density.
However, it is usually asserted that such a density cannot be
locally defined on the basis of the equivalence principle [1]. As a
consequence, a massive number of attempts have been made in GR
starting from Einstein who constructed his locally conserved
energy-momentum complex [2]. Following the same lines,
Landau-Lifshitz [3], Papapetrou [4], Bergman [5], Tolman [6] and
Weinberg [7] constructed their own energy-momentum which are also
locally conserved. The major drawback of these complexes is that one
can get acceptable results only by using Cartesian coordinate
system. M$\ddot{o}$ller removed this drawback and proposed a new
energy-momentum complex [8], which is not coordinate dependent.

In spite of some doubts [1], different authors [9-12] continue work
to explore this problem further. In particular, the idea of
quasi-local mass seems to be more clear [13]. According to this
approach, a Hamiltonian boundary term is associated to each
gravitational energy-momentum pseudo-tensor and consequently
energy-momentum complexes are quasi-local. One can use any
coordinate system in this formalism.

Many authors [14-20] defined energy of the gravitational field by
introducing the Hamiltonian approach in the framework of Schwinger's
condition [21]. According to this formalism, energy is given by an
integral of a scalar density in the form of total divergence which
appears as the Hamiltonian constraint of this theory. Andrade et al.
[15,22-24] constructed Lagrangian density of teleparallel equivalent
of GR, which provides an alternative geometrical framework for the
Einstein field equations (EFEs).

Virbhadra et al. [25-29] explored the energy-momentum distribution
of several spacetimes, such as, Kerr-Newmann, Kerr-Schild classes,
Einstein-Rosen, Vaidya and Bonnor-Vaidya spacetimes. They conclude
that different energy-momentum prescriptions provide the same
results which agree with those obtained by Penrose [30] and Tod [31]
in the framework of quasi-local mass. On the other hand, it is found
[28-29, 32-42] that M$\ddot{o}$ller prescription yields different
results from the other prescriptions by considering particular
examples including Schwarzschild spacetime.

Teleparallel gravity is an alternate description of gravitation
which corresponds to a gauge theory for the translation group [23].
The energy localization problem was re-considered in the framework
of this theory by many authors [43-49]. They showed that energy may
be localized in this formalism and found results which are quite
consistent with the available results in the framework of GR. Vargas
[45] found that the total energy of the closed
Friedmann-Robertson-Walker spacetime is zero by using teleparallel
version of Einstein and Landau-Lifshitz complexes. This exactly
coincides with that obtained by Rosen [50] in the framework of GR.

According to Lessner [51], M$\ddot{o}$ller prescription is a
powerful concept of energy and momentum in GR and also some authors
[39-43, 52-56] concluded that it is a good tool for evaluating
energy distribution in a given spacetime. The use of M$\ddot{o}$ller
prescription thus seems to be more interesting, useful and
appropriate while finding the energy distribution. In this paper, we
use teleparallel version of M$\ddot{o}$ller prescription and find
the components of energy-momentum densities of the Lewis-Papapetrou
spacetimes.

The paper is organized as follows. Section \textbf{2} will provide
some basic concepts of teleparallel theory of gravity and the
M$\ddot{o}$ller energy formalism in tetrad theory. Section
\textbf{3} is devoted to evaluate the components of the
energy-momentum densities of Lewis-Papapetrou spacetimes. The last
section provides summary and discussion of the results obtained.

\section{Teleparallel Gravity}

The theory of teleparallel gravity (TPG) is described by the
Weitzenb$\ddot{o}$ck connection given by [23]
\begin{eqnarray}
{\Gamma^\theta}_{\mu\nu}={{h_a}^\theta}\partial_\nu{h^a}_\mu,
\end{eqnarray}
where the non-trivial tetrad ${h^a}_\mu$ and its inverse field
${h_a}^\nu$ satisfy the relations
\begin{eqnarray}
{h^a}_\mu{h_a}^\nu={\delta_\mu}^\nu; \quad\
{h^a}_\mu{h_b}^\mu={\delta^a}_b.
\end{eqnarray}
We shall use the Latin alphabet $(a,b,c,...=0,1,2,3)$ to denote the
tangent space indices and the Greek alphabet
$(\mu,\nu,\rho,...=0,1,2,3)$ to denote the spacetime indices. The
Riemannian metric in TPG arises as a by product [23] of the tetrad
field given by
\begin{equation}
g_{\mu\nu}=\eta_{ab}{h^a}_\mu{h^b}_\nu,
\end{equation}
where $\eta_{ab}$ is the Minkowski spacetime such that
$\eta_{ab}=diag(+1,-1,-1,-1)$. In TPG, the gravitation is attributed
to torsion [57], which plays the role of force here. For the
Weitzenb$\ddot{o}$ck spacetime, the torsion is defined as [23]
\begin{equation}
{T^\theta}_{\mu\nu}={\Gamma^\theta}_{\nu\mu}-{\Gamma^\theta}_{\mu\nu},
\end{equation}
which is antisymmetric in nature. Due to the requirement of
absolute parallelism, the curvature of the Weitzenb$\ddot{o}$ck
connection vanishes identically. The Weitzenb$\ddot{o}$ck
connection also satisfies the relation given by
\begin{equation}
{{\Gamma^{0}}^\theta}_{\mu\nu}={\Gamma^\theta}_{\mu\nu}
-{K^{\theta}}_{\mu\nu},
\end{equation}
where
\begin{equation}
{K^\theta}_{\mu\nu}=\frac{1}{2}[{{T_\mu}^\theta}_\nu+{{T_\nu}^
\theta}_\mu-{T^\theta}_{\mu\nu}],
\end{equation}
is the {\bf contortion tensor} and ${{\Gamma^{0}}^\theta}_{\mu\nu}
$ are the Christoffel symbols.

Mikhail et al. [43] defined the superpotential (which is
antisymmetric in its last two indices) of the M$\ddot{o}$ller tetrad
theory as
\begin{equation}
{U_\mu}^{\nu\beta}=\frac{\sqrt{-g}}{2\kappa}P_{\chi\rho\sigma}^{\tau\nu\beta}
[{\Phi^\rho}g^{\sigma\chi} g_{\mu\tau}-\lambda g_{\tau\mu}
K^{\chi\rho\sigma}-(1-2\lambda) K^{\sigma\rho\chi}],
\end{equation}
where
\begin{equation}
P_{\chi\rho\sigma}^{\tau\nu\beta}= {\delta_\chi}^{\tau}
g_{\rho\sigma}^{\nu\beta}+{\delta_\rho}^{\tau}
g_{\sigma\chi}^{\nu\beta}-{\delta_\sigma}^{\tau}
g_{\chi\rho}^{\nu\beta},
\end{equation}
and $ g_{\rho\sigma}^{\nu\beta}$ is a tensor quantity defined by
\begin{equation}
g_{\rho\sigma}^{\nu\beta}={\delta_\rho}^{\nu}{\delta_\sigma}^{\beta}-
{\delta_\sigma}^{\nu}{\delta_\rho}^{\beta}.
\end{equation}
$K^{\sigma\rho\chi}$ is contortion tensor given by Eq.(6), $g$ is
the determinant of the metric tensor $g_{\mu\nu}$, $\lambda$ is
free dimensionless coupling constant of teleparallel gravity,
$\kappa$ is the Einstein constant and $\Phi_\mu$ is the basic
vector field given by
\begin{equation}
\Phi_\mu={K^\nu}_{\nu\mu}.
\end{equation}
We define the energy momentum density as
\begin{equation}
\Xi_\mu^\nu= U_\mu^{\nu\rho},_\rho,
\end{equation}
where comma means ordinary differentiation. The momentum 4-vector of
M$\ddot{o}$ller prescription can be expressed as
\begin{equation}
P_\mu ={\int}_\Sigma {\Xi_\mu^0} dxdydz,
\end{equation}
where  $P_0$ gives the energy and $P_1$, $P_2$ and $P_3$  are the
momentum components while the integration is taken over the
hyper-surface element $\Sigma$, which is described by
$x^0=t=constant$. The energy may be given in the form of surface
integral [8] as
\begin{equation}
E=\lim_{r \rightarrow \infty} {\int}_{{r=constant}}
{U_0}^{0\rho}u_\rho dS,
\end{equation}
where $u_\rho$ is the unit three-vector normal to the surface
element $dS$.

\section{Energy-Momentum Densities of Lewis-\\Papapetrou
Spacetimes}

Due to the non-linearity of EFEs, frame dragging effect and the
presence of black hole, the study of rotating black holes with
external source has been a favorite research topic of many
scientists. The gravitational effect of the external source can be
studied appropriately in the framework of stationary axisymmetric
spacetimes. Rotation is a universal phenomenon which seems to be
shared by all objects, at all possible scales [58]. All rotating
regular solutions have been found within the usual Lewis-Papapetrou
ansatz [59] for a stationary axially symmetric spacetimes with two
Killing vector fields $\partial_t$ and $\partial_\theta$. The class
of stationary axisymmetric solutions of the EFEs has been studied by
Galtsov [60] to include the gravitational effect of an {\it
external} source. On the other hand, such spacetimes are also
astrophysically important because they describe the exterior of a
body in equilibrium [61]. The Lewis-Papapetrou metric is given by
\begin{equation}
ds^2=e^{2\psi}(dt-\omega d\theta)^2-
e^{2(\gamma-\psi)}(d\rho^2+dz^2)- \rho^2e^{-2\psi}d\theta^2,
\end{equation}
where $\omega$ is the angular velocity and $\gamma, \psi, \omega $
are arbitrary functions of $\rho$ and $z$ only. It is mentioned here
that this metric reduces to static axially symmetric spacetime [62]
for $\omega=0$ and satisfies the following constraints
\begin{eqnarray}
\ddot{\psi}+ \frac{1}{\rho}\dot{\psi}+ \psi''=o,\\
\dot{\gamma}=\rho({\dot{\psi}}^2 - {\psi'}^2), \quad\quad
\gamma'=2\rho\dot{\psi}\psi',
\end{eqnarray}
where dot and prime denote the derivatives w.r.t. $\rho$ and $z$
respectively. Following the same procedure as in [63, 24], one can
find the tetrad components for the Lewis-Papapetrou metric as
follows
\begin{equation}
{h^a}_\mu=\left\lbrack\matrix {e^\psi &&& 0 &&& -\omega e^\psi &&& 0
\cr 0 &&& e^{\gamma-\psi}cos\theta &&& -\rho e^{-\psi} sin\theta &&&
0 \cr 0 &&& e^{\gamma-\psi} \sin \theta &&& \rho e^{-\psi} cos\theta
&&& 0 \cr 0 &&& 0 &&& 0 &&& e^{\gamma-\psi} \cr } \right\rbrack
\end{equation}
with its inverse
\begin{equation} {h_a}^\mu=\left\lbrack\matrix
{e^{-\psi} && 0 && 0 && 0 \cr -\omega\rho^{-1}e^{\psi} sin\theta
&& e^{-\gamma+\psi} cos\theta && -\rho^{-1} e^\psi sin\theta   &&
0 \cr \omega\rho^{-1}e^\psi cos\theta && e^{-\gamma+\psi}sin\theta
&& \rho^{-1}e^{\psi} cos\theta && 0 \cr 0        &&   0   &&  0
&& e^{-\gamma+\psi} \cr } \right\rbrack.
\end{equation}
We see that Eqs.(2) and (3) can be easily verified with the help
of Eqs.(17) and (18). Using Eqs.(17) and (18) in Eq.(1), we obtain
the following non-vanishing components of the Weitzenb$\ddot{o}$ck
connection
\begin{eqnarray}
{\Gamma^0}_{01}&=&\dot{\psi},\quad\ {\Gamma^0}_{03}=\psi',
\quad\ {\Gamma^0}_{12}=\omega \rho^{-1} e^\gamma, \nonumber\\
{\Gamma^0}_{21}&=&\omega \rho^{-1}-(\dot{\omega}
+2\omega\dot{\psi}), \quad\, {\Gamma^0}_{23}=-(\omega'+2\omega
\psi'),\nonumber\\
{\Gamma^1}_{11}&=&\dot{\gamma}-\dot{\psi}={\Gamma^3}_{31},\quad\
{\Gamma^1}_{22}=-\rho e^{-\gamma}, \quad\,
{\Gamma^1}_{13}=\gamma' -\psi'={\Gamma^3}_{33}, \nonumber\\
{\Gamma^2}_{12}&=&\rho^{-1} e^\gamma, \quad
{\Gamma^2}_{21}=\rho^{-1}(1-\rho\dot{\psi}),\quad\
{\Gamma^2}_{23}=-\psi'.
\end{eqnarray}
The corresponding non-vanishing components of the torsion tensor are
obtained by using Eq.(19) in Eq.(4). These are given by
\begin{eqnarray}
{T^0}_{01}&=&-\dot{\psi}=-{T^0}_{10}, \quad\
{T^0}_{03}=-\psi'=-{T^0}_{30},\nonumber\\
{T^0}_{12}&=&\omega\rho^{-1}(1-e^\gamma)-(\dot{\omega}+
2\omega\dot{\psi})=-{T^0}_{21},\nonumber\\
{T^0}_{23}&=&\omega'+2\omega\psi'= -{T^0}_{32},\quad
{T^1}_{13}=-\gamma'+\psi'=-{T^1}_{31}, \nonumber\\
{T^2}_{12}&=&\rho^{-1}(1-e^\gamma)-\dot{\psi} =-{T^2}_{21}, \quad
{T^2}_{23}=\psi'=-{T^2}_{32}, \nonumber\\
{T^3}_{31}&=&-\dot{\gamma}+\dot{\psi}=-{T^3}_{13}.
\end{eqnarray}
Substituting these values in Eq.(10 ), we obtain the following
non-vanishing components of the basic vector field
\begin{eqnarray}
\phi_1&=&\dot{\psi}-\dot{\gamma}-\rho^{-1}(1-e^\gamma), \\
\phi_3&=&\psi'- \gamma',
\end{eqnarray}
which can also be written as
\begin{eqnarray}
\phi^1&=& e^{2(\psi-\gamma)}\{\dot{\gamma}+\rho^{-1}(1-e^\gamma)-\dot{\psi}\}, \\
\phi^3&=& e^{2(\psi-\gamma)}(\gamma'-\psi').
\end{eqnarray}
Using Eq.(20) in Eq.(6), it yields the following non-zero components
of the contorsion tensor
\begin{eqnarray}
K^{010}&=& \frac{\dot{\psi}}{\rho^2}\{\rho^2e^{-2\gamma}-\omega^2
e^{2(2\psi-\gamma)}\},\quad K^{030}=
\frac{\psi'}{\rho^2}\{\rho^2e^{-2\gamma}-\omega^2
e^{2(2\psi-\gamma)}\},\nonumber\\
K^{131}&=& e^{4(\psi-\gamma)}(\psi'-\gamma'), \quad K^{212}=
\frac{1}{\rho^3}(\rho\dot{\psi}+e^\gamma-1)
e^{2(2\psi-\gamma)},\nonumber\\
K^{232}&=& \frac{\psi'}{\rho^2} e^{2(2\psi-\gamma)}, \quad
K^{313}= e^{4(\psi-\gamma)(\dot{\psi}-\dot{\gamma})},\nonumber\\
K^{120}&=& K^{021}=K^{102}=
\frac{1}{2\rho^3}\{\omega(1-e^\gamma)-\rho(\dot{\omega}+2\omega\dot{\psi})\}
e^{2(2\psi-\gamma)},\nonumber\\
K^{320}&=&
K^{023}=K^{302}=\frac{-1}{2\rho^2}(\omega'+2\omega\psi')
e^{2(2\psi-\gamma)}.
\end{eqnarray}
It is worth mentioning here that the contorsion tensor is
antisymmetric w.r.t. its first two indices. Making use of
Eqs.(23)-(25) in Eq.(7), we get the required independent
non-vanishing components of the supperpotential in M$\ddot{o}$ller
tetrad theory as
\begin{eqnarray}
U_0^{01}&=&\frac{1}{\kappa}\{3\rho\dot{\psi}-2\rho \dot{\gamma} +
2 e^\gamma-2+\frac{\omega e^{4\psi}}{2 \rho^2}(\lambda-1)
(\rho\dot{\omega}+2\rho\omega \dot{\psi}-\omega+\omega
e^\gamma)\}, \nonumber\\
U_0^{03}&=&\frac{1}{\kappa}\{3\rho\psi'-2\rho\gamma'+ \frac{\omega
e^{4\psi}}{2\rho}(\lambda-1) (\omega'+2\omega \psi')\}, \nonumber\\
U_0^{12}&=&\frac{e^{4\psi}}{\rho^2\kappa}[\omega(\rho\dot{\psi}-e^\gamma-1)
+\frac{1}{2}(1+\lambda)\{\omega(1-e^\gamma)-\rho(\dot{\omega}+2
\omega\dot{\psi})\}], \nonumber\\
U_0^{23}&=&\frac{e^{4\psi}}{\rho
\kappa}[-\omega\psi'+\frac{1}{2}(1+\lambda)
(\omega'+2\omega\psi')], \nonumber\\
U_1^{02}&=&\frac{1}{2\rho^2\kappa}e^{2\gamma}(1-\lambda)[\omega(e^\gamma-1)+
\rho(\dot{\omega}+2\omega\dot{\psi})], \nonumber\\
U_2^{01}&=& \frac{1}{\kappa}[-\rho\omega\dot{\psi}+\frac{1}{\rho}
\omega^3\dot{\psi}e^{4\psi}+\frac{1}{2}(\lambda-1)\{\omega(e^\gamma-1)
+\rho(\dot{\omega}+2\omega \dot{\psi})\}], \nonumber\\
U_2^{03}&=&\frac{1}{\kappa}[-\rho\omega\psi'+\frac{1}{\rho}\omega^3
\psi'e^{{4\psi}+\frac{\rho}{2}(\lambda-1)(\omega'+2 \omega
\psi'})], \nonumber\\
U_3^{02}&=&\frac{1}{2\rho\kappa}e^{2\gamma}(1-\lambda)(\omega'+2
\omega\psi')].
\end{eqnarray}
Substituting these values in Eq.(11), it follows that the
energy-momentum density components
\begin{eqnarray}
\Xi_0^0&=&\frac{1}{\kappa}[3\rho(\ddot{\psi}+\psi'')-2\rho
(\ddot{\gamma}+\gamma'')+ 3\dot{\psi}+2
\dot{\gamma}(e^\gamma-1)+\frac{1}{2\rho^2} e^{4\psi}(\lambda-1)\nonumber\\
&&\{\rho(\dot{\omega}^2+{\omega'}^2)+8\rho\omega(\dot{\omega}\dot{\psi}+\omega'\psi')
+8\rho\omega^2(\dot{\psi}^2+\psi'^2)+\rho\omega(\ddot{\omega}+\omega'')\nonumber\\
&&+2\rho\omega^2(\ddot{\psi}+\psi'')-2\omega^2\dot{\psi}+2\rho\omega^2\ddot{\psi}-
3\omega\dot{\omega}+\frac{2{\omega}^2}{\rho}+\omega
e^\gamma(\omega \dot{\gamma}+2\dot{\omega}\nonumber\\
&&+4\omega\dot{\psi}-\frac{2\omega}{\rho})\} ],\nonumber\\
\Xi_2^0&=&\frac{1}{\kappa}[-\omega\dot{\psi}
-\rho(\dot{\omega}\dot{\psi}+\omega'\psi')
-\rho\omega(\ddot{\psi}+\psi'')
+\frac{\omega^2}{\rho}e^{4\psi}\{3(\dot{\omega}\dot{\psi}
+\omega'\psi')\nonumber\\&&+\omega(\ddot{\psi}+\psi'')
+4\omega(\dot{\psi}^2+\psi'^2)
\frac{\omega}{\rho}\dot{\psi}\}+\frac{1}{2}
(1+\lambda)\{e^\gamma(\dot{\omega}+
\omega\dot{\gamma})\nonumber\\
&&+2\rho(\dot{\omega}\dot{\psi}
+\omega'\psi')+2\rho\omega(\ddot{\psi}+\psi'')
+\rho(\ddot{\omega}+\omega'')+2\omega\ddot{\psi}\}].
\end{eqnarray}
We see that the energy-momentum turns out to be non-vanishing and
well-defined quantities. The component of the momentum density is
non-zero only along the $\theta$-direction which is due to the cross
term $dtd\theta$ involving in the given metric. If we take
$\omega=0$ and use Eqs.(15) and (16), the above results take the
following form
\begin{eqnarray}
\Xi_0^0&=&\frac{2}{\kappa}(2\rho {\psi'}^2+\dot{\gamma}e^\gamma),\nonumber\\
\Xi_i^0&=&0,\quad (i=1, 2, 3).
\end{eqnarray}
It is worth mentioning that these results are exactly the same as
found by us in the case of static axisymmetric spacetimes [64].

\section{Summary and Discussion}

The debate of localization of energy-momentum has been an open issue
since the time of Einstein when he formulated the well-known
relation between mass and energy. Misner et al. [2] concluded that
energy can only be localized in spherical coordinates. But, soon
after, Cooperstock and Sarracino [65] demonstrated that if the
energy is localizable in spherical systems then it can be localized
in any system. Bondi [66] rejected the idea of non-localization of
energy in GR due to the reason that there should be any form of
energy which contributes to gravitation and hence its location can,
in principle, be found. Many authors believed that a tetrad theory
should describe more than a pure gravitation field [67]. In fact,
M$\ddot{o}$ller [68] considered this possibility in his earlier
attempt to modify GR. Recently, Salti et. al. [46-48, 69-70]
explored many examples in the context of both GR and TPT and found
consistent results.

In this work, we have evaluated the energy-momentum density
components of stationary axisymmetric Lewis-Papapetrou spacetimes.
For this purpose, we use the teleparallel version of M$\ddot{o}$ller
prescription. It is found that the energy density of the
Lewis-Papapetrou spacetime turns out to be non-zero and well-defined
and the momentum becomes constant except along $\theta$ direction.
This is due to the fact that the Lewis-Papapetrou metric contains a
cross term $dtd\theta$. Further, we note that for $\omega=0$, these
results exactly match with those of the Weyl metrics in the context
of teleparallel theory of gravity [64]. It is mentioned here that
our results of TPT do not coincide with those of GR.

\vspace{0.5cm}

{\bf Acknowledgment}

\vspace{0.5cm}

We acknowledge the enabling role of the Higher Education
Commission Islamabad, Pakistan, and appreciate its financial
support through the {\it Indigenous PhD 5000 Fellowship Program
Batch-I}.
%\newpage
\vspace{0.5cm}

{\bf References}

\begin{description}

\item{[1]} Misner, C.W., Thorne, K.S. and Wheeler, J.A.:
           \textit{Gravitation} (Freeman, New York, 1973).

\item{[2]} Trautman, A.: \textit{Gravitation}: \textit{An Introduction to Current
           Research}, ed. by Witten, L. (Wiley, New York, 1962).

\item{[3]} Landau, L.D. and Lifshitz, E.M.: \textit{The Classical Theory
           of Fields} (Addison-Wesley Press, New York, 1962).

\item{[4]} Papapetrou, A.: \textit{Proc. R. Irish Acad. } \textbf{A52}(1948)11.

\item{[5]} Bergman, P.G. and Thomson, R.: Phys. Rev.
           \textbf{89}(1958)400.

\item{[6]} Tolman, R.C.: \textit{Relativity, Thermodynamics and
           Cosmology} (Oxford University Press, Oxford, 1934).

\item{[7]} Weinberg, S.: \textit{Gravitation and Cosmology} (Wiley, New
           York, 1972).

\item{[8]} M$\ddot{o}$ller, C.: Ann. Phys. (N.Y.) \textbf{4}(1958)347.

\item{[9]} Dubois-Violette, M. and Madore, J.: Math Phys.
\textbf{108}(1987)213.

\item{[10]} Szabados, B.: Class. Quant. Gravit. \textbf{9}(1992)2521.

\item{[11]} Aguirregabiria, J.M., Chamorro, A. and Virbhadra, K.S.: Gen.
           Relativ. Gravit. \textbf{28}(1996)1393.

\item{[12]} Tshirafuji, T. and Nashed, G.L.: Prog. Theor. Phys. \textbf{98}(1997)1355.

\item{[13]} Chang, C.C. and Nester, J.M.: Phys. Rev. Lett.
\textbf{83}(1999)1897 and references therein.

\item{[14]} Maluf, J.W.: J. Math. Phys. \textbf{35}(1994)335.

\item{[15]} De Andrade, V.L. and Pereira,  J.G.: Phys. Rev. {\bf
            D56}(1997)4689.

\item{[16]} Blagojevic, M. and Nikolic, I.A.: \textbf{D62}(2000)024021.

\item{[17]} Mielke, E.W.: Ann. Phys. (NY) \textbf{219}(1992)78.

\item{[18]} Mielke, E.W.: Phys. Lett. \textbf{A 251}(1999)349.

\item{[19]} Hehl, F.W.: \textit{Proceeding of 8th Marcel Grossman meeting,
            Jerusalem, 1997, Tsvi Piran ed., World Scientific, Singapore London,
            1998}

\item{[20]} Maluf, J.W.: J. Math. Phys. \textbf{36}(1995)4242.

\item{[21]} Schwinger, J.: Phys. Rev. \textbf{130}(1963)1253.

\item{[22]} De Andrade, V.L, Guillen, L.C.T and Pereira, J.G.: Phys. Rev. Lett.
            {\bf 84}(2000)4533.

\item{[23]} Aldrovandi, R. and Pereira, J.G.: {\it An Introduction to
            Gravitation Theory} (preprint)

\item{[24]} Hayashi, K. and Tshirafuji: Phys. Rev. {\bf D19}(1979)3524.

\item{[25]} Virbhadra, K.S.: Phys. Rev. \textbf{D60}(1999)104041.

\item{[26]} Virbhadra, K.S.: Phys. Rev. \textbf{D42}(1990)2919.

\item{[27]} Virbhadra, K.S.: Phys. Lett. \textbf{B331}(1994)302.

\item{[28]} Virbhadra, K.S. and Parikh, J.C.: Phys. Lett.  \textbf{B317}(1993)312

\item{[29]} Rosen, N. and Virbhadra, K.S.: Gen. Gravit.  \textbf{25}(1993)429.

\item{[30]} Penrose, R.: \textit{Proc. Roy. Soc., London }\textbf{A381}(1982)53.

\item{[31]} Tod, K.P.: \textit{Proc. Roy. Soc., London }\textbf{A388}(1983)457.

\item{[32]} Xulu, S.S.: Astrophys. Space Sci. \textbf{283}(2003)23.

\item{[33]} Xulu, S.S.: Mod. Phys. Lett. {\bf A15}(2000)1511.

\item{[34]} Sharif, M.: Int. J. Mod. Phys. \textbf{A17}(2002)1175.

\item{[35]} Sharif, M.: Int. J. Mod. Phys. \textbf{A18}(2003)4361.

\item{[36]} Sharif, M.: Int. J. Mod. Phys. \textbf{A19}(2004)1495.;

\item{[37]} Sharif, M.: Int. J. Mod. Phys.  \textbf{D13}(2004)1019.

\item{[38]} Sharif, M. and Fatima, T.: Nouvo Cim. \textbf{B120}(2005)533.

\item{[39]} Radinschi, I.: Mod. Phys. Lett. \textbf{A16}(2001)673.

\item{[40]} Halpern, P.: gr-qc/0606095.

\item{[41]} Yang, I-Ching and Radinschi, I.: Chin. J. Phys. {\bf
42}(2004)40.

\item{[42]} Yang, I-Ching and Radinschi, I.: Chin. J. Phys. {\bf 41}(2003)326.

\item{[43]} Mikhail, F.I., Wanas, M.I., Hindawi, A. and Lashin, E.I.: Int. J. Theo.
            Phys. \textbf{32}(1993)1627.

\item{[44]} Nashed, G.G.L.: Phys. Rev. \textbf{D66}(2002)060415.

\item{[45]} Vargas, T.: Gen. Rel. Grav. \textbf{30}(2004)1255.

\item{[46]} Salti, M., Havare, A.: Int. J. of Mod. Phys.
\textbf{A20}(2005)2169.

\item{[47]} Salti, M.: Int. J. of Mod. Phys. \textbf{A20}(2005)2175.

\item{[48]} Salti, M.: Astrophys. Space Sci. \textbf{229}(2005)159.

\item{[49]} Aydogdu, O. and Salti, M.: Astrophys. Space Sci.  \textbf{229}(2005)227.

\item{[50]} Rosen, N.: Gen. Rel. Grav. \textbf{26}(1994)323.

\item{[51]} Lessner, G.: Gen. Rel. Grav.  \textbf{28}(1996)527.

\item{[52]} Gad, R.M.: Astrophys. Space Sci. \textbf{295}(2004)495.

\item{[53]} Vagenas, E.C.: Int. J. Mod. Phys.
\textbf{A18}(2003)5781.

\item{[54]} Vagenas, E.C.: Int. J. Mod. Phys. \textbf{A18}(2003)5949.

\item{[55]} Vagenas, E.C.: Mod. Phys. Lett. \textbf{A19}(2003)213.

\item{[56]} Vagenas, E.C.: arXiv:gr-qc/0602107.

\item{[57]} Aldrovandi, R. and Pereira, J.G.: {\it An Introduction to
            Geometrical Physics} (World Scientific, 1995).

\item{[58]} Radu, E.: arXiv:gr-qc/0512094.

\item{[59]} Wald, R.: {\it General Relativity} (University of Chicago Press, Chicago, 1984).

\item{[60]} Galtsov, D.V.: arXiv:gr-qc/9808002.

\item{[61]} Islarn, J.N.: \textit{Rotating Fields in General Relativity}
           (Cambridge Univ. Press, Cambridge, 1985).

\item{[62]} Curzon, H.E.J.: \textit{Proc. Math. Soc. London,} {\bf 23}(1924)477.

\item{[63]} Pereira, J.G., Vargas, T. and Zhang, C.M.: Class. Quant. Grav.
            {\bf 18}(2001)833.

\item{[64]} Sharif, M. and Amir, M.J.: Submitted for publication.

\item{[65]} Cooperstock, F.I. and Sarracino, R.S.: J. Phys. A: Math.
            Gen. \textbf{11}(1978)877.

\item{[66]} Bondi, H.: \textit{ Proc. R. Soc. London } \textbf{A427}(1990)249.

\item{[67]} Nashed, G.G.L.: Nouvo Cim. \textbf{B117}(2002)521.

\item{[68]} M$\ddot{o}$ller, C.: Mat. Fys. Medd. Dan. Vid. Selsk. \textbf{1}(1961)10.

\item{[69]} Aydogdu, O. and Salti, M.: Astrophys. Space Sci.
\textbf{229}(2005)227.

\item{[70]} Aydogdu, O., Salti, M. and Korunur, M.: Acta Phys. Slov.
            \textbf{55}(2005)537.

\end{description}
\end{document}